\documentclass[runningheads]{llncs}
\usepackage[T1]{fontenc}
\usepackage{graphicx}
\usepackage{booktabs}
\usepackage[misc]{ifsym}

\usepackage{mwe}
\usepackage{graphicx}
\usepackage{booktabs}
\usepackage{xcolor}
\usepackage{caption}
\usepackage{subcaption}
\usepackage{multirow}
\usepackage{siunitx}
\usepackage{tikz}
\usepackage{zebra-goodies}
\usepackage{amsmath,amssymb,amsfonts}

\begin{document}

\title{PRIMS: \underline{P}hysics-guided \underline{R}epresentation for Fluid \underline{I}dentification in \underline{M}ultimodal \underline{S}ensing}
\titlerunning{PRIMS: Physics-guided Representation for Fluid Identification}

\toctitle{PRIMS: \underline{P}hysics-guided \underline{R}epresentation for Fluid \underline{I}dentification in \underline{M}ultimodal \underline{S}ensing}


\tocauthor{Hai-Long Nguyen, Trung Thanh Nguyen, Lars Holm, Dennis Alveringh, Duc V. Le}

\author{
Hai-Long Nguyen\inst{1} (\Letter) \and
Trung Thanh Nguyen\inst{2} \and
Lars Holm\inst{1} \and \\
Dennis Alveringh\inst{1} \and
Duc V. Le\inst{1}
}

\authorrunning{H.-L. Nguyen et al.}



\institute{
University of Twente, Netherlands\\
\email{\{hailong.nguyen,l.holm,d.alveringh,v.d.le\}@utwente.nl}
\and
Nagoya University, Japan\\
\email{nguyent@cs.is.i.nagoya-u.ac.jp}
}

\maketitle              
\begin{abstract}
Accurate on-device fluid identification is essential for microfluidic applications, yet maintaining reliability under varying flow, pressure, and temperature remains a key challenge. Existing learning-based methods often treat sensor signals as  domain-agnostic features, neglecting the underlying physical relationships that govern fluid behavior, thereby limiting generalization and interpretability. To address this, we propose \textbf{PRIMS}, a physics-aware multimodal Transformer that integrates physical knowledge into representation learning and attention mechanisms through three dedicated modules: (1) \textit{Physics-based Token Vectorization} transforms raw Coriolis and pressure sensor signals into physically meaningful token embeddings; (2) \textit{Physical Component Synthesizer} models viscosity-related dependencies among flow, pressure, and density; and (3) \textit{Physics-guided Fusion} captures cross-physical correlations through attention-based integration. By embedding these physics-based relationships directly into the model architecture, PRIMS bridges analytical fluid mechanics and deep learning, enabling interpretable, data-efficient, and resilient fluid classification. Evaluations on a five-fluid benchmark under dynamic flow, pressure, and temperature conditions show that PRIMS achieves 98.92\% average F1-score with only 0.46 million parameters, a 14$\times$ reduction compared to state-of-the-art Transformer-based methods. PRIMS also consistently outperforms prior SOTA models under out-of-distribution shifts to unseen temperature ranges and unseen flow-rate ranges, indicating strong robustness to operating conditions not observed during training. These findings suggest that designing architectures that explicitly mirror governing physical relationships can make them learn transferable, environment-independent representations, improving real-world reliability for microfluidic sensing.


\end{abstract}

\section{Introduction}
\label{sec:introduction}

Microfluidic sensing technologies have become central to a wide range of industrial, 
biomedical, and environmental applications~\cite{ejeian2019design}. 
Over the last decades, advances in sensor miniaturization and integration have enabled the simultaneous measurement of multiple flow parameters on a single 
chip~\cite{van1995multi,park2024machine}.
Among these, Coriolis mass-flow sensors co-integrated with differential pressure sensors 
provide a rich, multimodal description of fluid behavior~\cite{schut2018fully}. 
The Coriolis sensor directly measures mass flow and density through tube oscillations, 
while the pressure sensors capture viscosity-dependent pressure drops, 
together enabling physical characterization of the fluid. 

Despite these advances, conventional analytical models for fluid characterization 
rely on idealized assumptions that often fail under nonlinear or variable operating regimes, such as temperature fluctuations, turbulence, or low-flow conditions. 
To overcome these limitations, learning-based methods have been introduced to extract features directly from time-series sensor signals~\cite{ALVERINGH2023114762,zubavicius2024fluid}. 
Convolutional and recurrent architectures~\cite{ALVERINGH2023114762} have demonstrated the ability to classify fluids and infer rheological parameters, offering data-driven adaptability beyond static analytical formulas. 
However, these models typically treat sensor readings as domain-agnostic features, neglecting the underlying physical dependencies among flow rate, pressure, and density that govern the system's behavior. 
As a result, their predictions often degrade when operating conditions deviate from those seen during training, limiting generalization and robustness.

Recent developments in physics-informed learning~\cite{raissi_jcp_2019,hu2025physics,pham2026padm} highlight the potential of integrating physical constraints within neural architectures to enhance interpretability and stability. 
For flow systems, physical laws such as the Hagen--Poiseuille equation~\cite{kirby_cup_2010} establish deterministic relations among measurable quantities like pressure drop, viscosity, and channel geometry. 
However, existing physics-informed approaches have not been adapted to multimodal microfluidic sensing, where multiple coupled physical quantities must be jointly modeled.
Meanwhile, learning-based methods such as iTransformer~\cite{ICLR2024_2ea18fdc}, with approximately 6.50 million trainable parameters, have demonstrated strong capability in modeling temporal dependencies across multimodal sensor streams. 
Yet their large model size and domain-agnostic design limit practicality for embedded or edge devices, where computational resources and power budgets are constrained. 
These observations highlight the need for compact, physics-aware architectures that integrate domain knowledge directly into the learning process.

To address these limitations, we propose \textbf{PRIMS}, a physics-aware multimodal Transformer designed for fluid classification in microfluidic systems. 
PRIMS integrates physical principles directly into the learning process through three modules. 
(1) \textit{Physics-based Token Vectorization} encodes raw Coriolis and pressure signals into compact and physically meaningful representations, motivated by the observation that spectral and differential features directly relate to mass flow, density, and pressure drop. 
(2) \textit{Physical Component Synthesizer} models intrinsic relationships among mass flow, density, and pressure through cross-attention, reflecting the viscosity coupling described by the Hagen--Poiseuille law. 
(3) \textit{Physics-guided Fusion} integrates these multimodal representations through self-attention, enabling the model to capture cross-physical dependencies while maintaining interpretability. 
By embedding physical structure into the Transformer architecture, PRIMS bridges the gap between analytical modeling and data-driven learning, achieving interpretable and resilient fluid classification under varying flow, temperature, and pressure conditions.

The main contributions are as follows:
\begin{itemize}
    \item We propose PRIMS, a physics-aware multimodal Transformer 
    that embeds physical relationships among flow, pressure, and density 
    directly into its tokenization and attention layers, enabling 
    efficient and interpretable deployment on edge devices.
    
    \item We design three physics-aware modules, namely 
    Physics-based Token Vectorization, Physical Component Synthesizer, 
    and Physics-guided Fusion, that jointly transform raw sensor signals 
    into physics-consistent representations, synthesize viscosity-related 
    features, and fuse multimodal information to capture cross-physical 
    relationships.
    

    \item We provide a comprehensive experimental evaluation across five fluids under varying flow and thermal conditions, including out-of-distribution shifts, demonstrating that PRIMS consistently achieves the best F1-score across all experimental settings while using 14$\times$ fewer parameters than state-of-the-art Transformer-based methods.
\end{itemize}

\section{Related Work}
\label{sec:related}

\noindent \textbf{Learning-based Fluid Identification.}
Low-flow sensing in microfluidic systems is commonly realized through thermal, differential-pressure, ultrasonic, and Coriolis principles~\cite{4675306,haneveld_jmm_2010}.
Recent research has increasingly integrated machine learning with microfluidic sensing to interpret complex, multivariate signals and infer fluid properties directly from sensor data.
Alveringh et al.~\cite{ALVERINGH2023114762} propose a micro-Coriolis--pressure fusion framework using a deep learning model to classify multiple fluids from combined flow and pressure signals, demonstrating the promise of learning-based fluid identification.
Related efforts have applied machine learning to infer rheological properties such as viscosity from microchannel deflection data~\cite{mustafa2023machine}, and to detect complex biological mixtures through integrated microfluidic-sensor platforms~\cite{yang2025integrated}. 
Despite these advances, current learning-based methods often treat sensor outputs as independent features and exhibit sensitivity to domain shifts in viscosity, density, and flow regime, with limited incorporation of physical fluid relations. 

\vspace{5pt}
\noindent \textbf{Physics-guided Learning for Flow Systems.}
Integrating physical constraints into learning, through Physics-Informed Neural Networks (PINNs) or differentiable Partial Differential Equation (PDE) solvers, has been shown to improve extrapolation under domain shifts~\cite{raissi_jcp_2019}. 
In microfluidic systems, analytic relations such as the Hagen--Poiseuille equation link pressure drop, viscosity, and geometry in laminar 
regimes~\cite{kirby_cup_2010}. 
In mechanical and chemical sensing, cross-modal attention mechanisms have achieved state-of-the-art performance by learning interaction priors between physical variables~\cite{yang_entropy_2022,koupai2022self}, and physics-aware fusion networks have been explored to align heterogeneous sensor modalities within shared latent spaces.
However, these physics-guided methods have largely focused on mechanical or chemical domains and have not been adapted to multimodal microfluidic sensing, where coupled quantities such as flow, pressure, and density must be jointly modeled.

\vspace{5pt}
\noindent \textbf{Efficient and Transformer-based Sensing Models.}
Transformers~\cite{vaswani2017attention} have rapidly become a foundation for multivariate time-series and sensor fusion tasks~\cite{nguyen2025multisensor}, benefiting from their flexible attention structure.
Models such as Informer~\cite{zhou2021informer} and Autoformer~\cite{wu2021autoformer} introduced efficient sparse attention 
for long-horizon forecasting, while more recent work like iMoT~\cite{nguyen2025imot} adapts attention mechanisms for physical sensing by modeling temporal alignment and uncertainty across motion modalities.
In parallel, compact Transformer variants such as EfficientFormer~\cite{li2022efficientformer} and EdgeViT~\cite{edgevit_eccv_2022} demonstrate that attention-based architectures can achieve real-time inference on resource-constrained hardware.
These advances show that Transformers can jointly capture complex temporal dependencies and remain computationally practical, yet most existing designs focus on motion or vision tasks rather than coupled physical processes in fluids.

\section{Preliminaries: Fluid Classification Theory}
\label{sec:fluid_classification_theory}

In this section, we present the fundamental physical properties that underlie fluid behavior in microfluidic systems. 
Traditional fluid identification relies on direct measurements of density, dynamic viscosity, and kinematic viscosity, which are strongly linked to molecular composition and temperature~\cite{cengelFundamentalsThermalFluidSciences2001}. 
Sensor calibration using reference fluids allows the construction of lookup tables or multidimensional property spaces for classification. 
As shown in Table~\ref{tab:fluids}, many liquids exhibit similar 
densities or viscosities, which further limits discriminative power. 
These limitations motivate the physics-aware learning approach 
introduced in Section~\ref{sec:proposed_method}.

\begin{table}[t]
\centering
\caption{
Physical properties of five fluids used in this study, based on Bronkhorst\textsuperscript{\textregistered} FLUIDA\textsuperscript{\textregistered}~\cite{bronkhorst_ml120v21_datasheet}. 
Values correspond to a single gauge-pressure condition.
}
\label{tab:fluids}
\small
\setlength{\tabcolsep}{4pt}
{
\begin{tabular}{llcccc}
\toprule
\textbf{Fluid} & \textbf{Abbr.} & \textbf{Phase} &
\textbf{Density} & \textbf{Kinematic Visc.} & \textbf{Dynamic Visc.} \\
 &  &  & (\si{\kilo\gram\per\meter\cubed}) & (\si{\milli\meter\squared\per\second}) & (\si{\milli\pascal\second}) \\
\midrule
Nitrogen    & N\textsubscript{2}  & Gas & 7.05 & 2.44 & 0.018 \\
Water       & H\textsubscript{2}O & Liq. & 999  & 1.16 & 1.15 \\
Isopropanol & IPA                 & Liq. & 800  & 3.40 & 2.72 \\
Ethanol     & EtOH                & Liq. & 812  & 1.53 & 1.27 \\
Acetone     & Ace                 & Liq. & 805  & 0.42 & 0.34 \\
\bottomrule
\end{tabular}
}
\footnotesize
\textit{\\Note:} Abbr. = Abbreviation; Liq. = Liquid; Visc. = Viscosity.
\vspace{-15pt}
\end{table}

\vspace{5pt}
\noindent \textbf{Phase.} 
In fluid mechanics, the term phase refers to the physical state of matter, which can exist as a solid, liquid, or gas. 
Each phase exhibits distinct thermodynamic and transport properties that govern flow behavior. 
Since solids do not undergo continuous deformation, microfluidic sensing typically focuses on liquids and gases, which can exhibit measurable flow characteristics.

\vspace{5pt}
\noindent \textbf{Density.} 
Denoted by $\rho = m / V$, is defined as the mass of a fluid $(m)$ per unit volume $(V)$~\cite{cengelFundamentalsThermalFluidSciences2001}. 
It governs buoyancy, pressure distribution, and the overall dynamics of fluid motion. 
Variations in density, particularly in compressible or thermally driven flows, can strongly influence flow behavior and the measurable response of microfluidic sensors.

\vspace{5pt}
\noindent \textbf{Dynamic Viscosity.} 
Denoted by $\eta = \tau \big/ \frac{du}{dy}$, quantifies a fluid's internal resistance to shear deformation~\cite{cengelFundamentalsThermalFluidSciences2001}, where $\tau$ is the shear stress and $\frac{du}{dy}$ is the velocity gradient perpendicular to the flow direction. 
For Newtonian fluids, $\eta$ remains constant under varying shear rates, while non-Newtonian fluids exhibit shear-dependent viscosity that requires more complex rheological modeling.

\vspace{5pt}
\noindent \textbf{Kinematic Viscosity.}
Denoted by $\nu = \eta / \rho$, represents the ratio 
of dynamic viscosity $(\eta)$ to density $(\rho)$~\cite{cengelFundamentalsThermalFluidSciences2001}.
This property describes the balance between viscous and inertial forces 
and determines flow regimes characterized by the Reynolds number, 
which governs the transition between laminar and turbulent behavior. 
Together, dynamic and kinematic viscosity define the energy dissipation 
and resistance to deformation in fluid motion.


\section{Methodology}
\label{sec:proposed_method}
\begin{figure}[t]
	\centering
    \includegraphics[width=0.75\columnwidth]{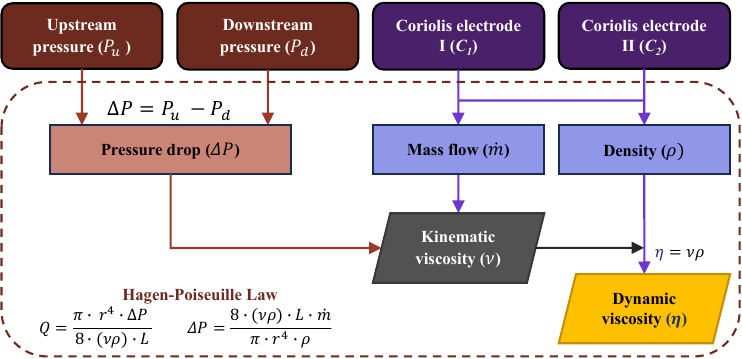}
	\caption{
    Principle of fluid classification using an integrated 
    Coriolis mass flow sensor with upstream and downstream pressure sensors. 
    The sensors directly measure pressure ($P_u$, $P_d$), mass flow ($\dot{m}$), and density ($\rho$), from which kinematic and dynamic viscosity ($\mathcal{V}$, $\eta$) are derived via the Hagen--Poiseuille law, where $L$ and $r$ denote the length and inner radius of the flow channel.
    }
	\label{fig:sensor}
\end{figure}

\noindent \textbf{Motivation.}
Among existing fluid sensing technologies, micro-fabricated Coriolis mass-flow sensors integrated with resistive pressure sensors introduced in~\cite{7994261} represent the state-of-the-art, offering direct and high-precision measurements of mass flow and density, and enabling viscosity estimation when combined with pressure sensors (Figure~\ref{fig:sensor}).
These sensors provide stability and compact integration, making them well suited for embedded microfluidic applications.
However, physics-based analytical approaches relying solely on these sensors struggle with nonlinear flow regimes, environmental variation, and overlapping fluid properties.
PRIMS addresses these challenges by embedding fluid-domain knowledge directly into a multimodal Transformer architecture.

\vspace{5pt}
\noindent \textbf{Problem Formulation.}
The objective of this study is to classify fluid types from time-series signals acquired by an integrated Coriolis mass-flow and pressure sensing system (Figure~\ref{fig:sensor}). 
The Coriolis sensor produces two sinusoidal signals, $C_1(t)$ and $C_2(t)$, whose phase difference $\Delta \phi$ is proportional to the mass flow rate, while the resonance frequency depends on the fluid density $\rho$~\cite{ALVERINGH2023114762}. 
Two resistive pressure sensors located upstream and downstream provide voltage signals $P_u(t)$ and $P_d(t)$, from which the pressure drop is obtained as $\Delta P = P_u - P_d$.
By combining the Coriolis and pressure modalities, the system enables simultaneous estimation of fluid properties, including density $\rho$, dynamic viscosity $\eta$, and kinematic viscosity $\nu$, which serve as discriminative features for fluid classification.

\begin{figure}[t]
\includegraphics[width=\textwidth]{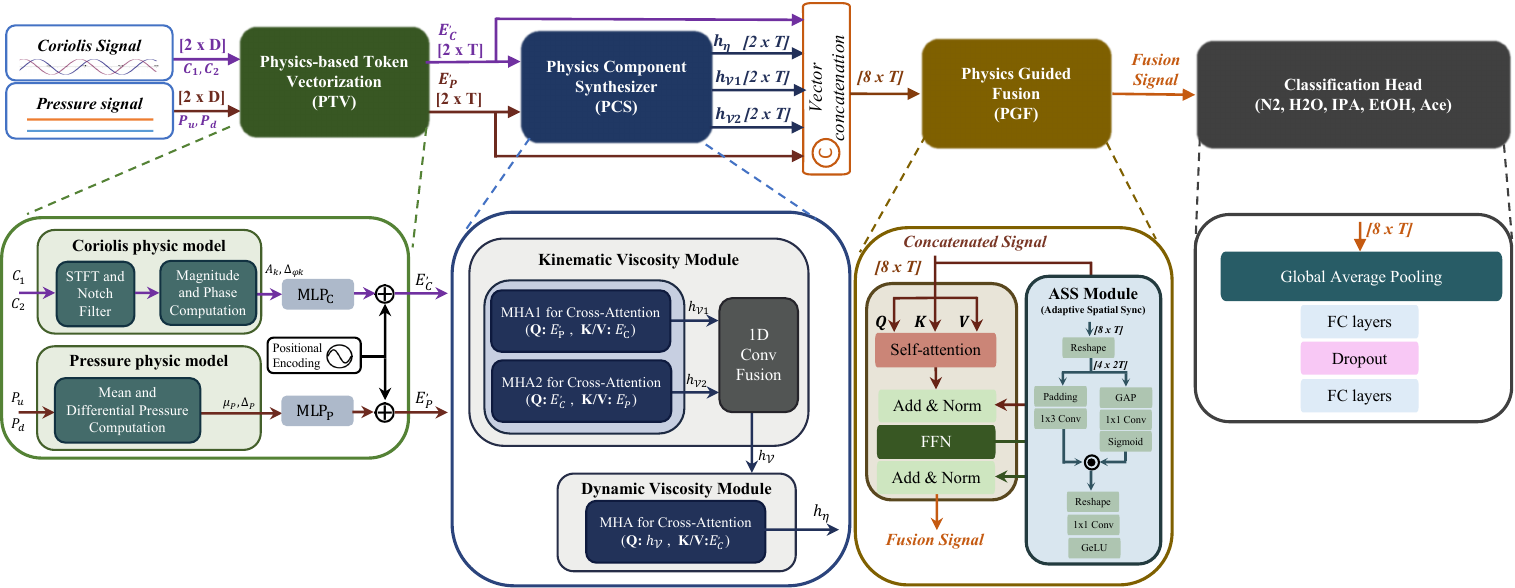}
\caption{
    Overview of the proposed PRIMS. The model consists of three components: (1) {Physics-based Token Vectorization (PTV)} 
    that transforms raw Coriolis and pressure signals into compact physics-informed representations; 
    (2) {Physical Component Synthesizer (PCS)} that models relationships among physical quantities through cross-attention-based viscosity modules; and 
    (3) {Physics-guided Fusion (PGF)} that integrates multimodal features via self-attention and adaptive spatial synchronization. 
    The fused representation is passed to a classification head for fluid-type prediction.
    Here $T$ denotes the embedding dimension $d_\text{embed}$.
    }
\label{fig:mosaic-physical-attentionbased}
\end{figure}

Each measurement window $\mathcal{X}^i$ from the Coriolis and pressure modalities spans a fixed temporal duration $\mathcal{T}$ (milliseconds), sampled at a frequency $F_s$ (samples per second), resulting in $\mathbf{D} = \frac{F_s}{1000}\mathcal{T}$ time steps.
The recorded data are represented as multivariate time series from two modalities as:
\begin{equation}
    \mathbf{X}_c = [C_1, C_2], \quad 
    \mathbf{X}_p = [P_u, P_d],
\end{equation}
where $\mathbf{X}_c, \mathbf{X}_p \in \mathbb{R}^{2 \times \mathbf{D}}$. 
Each training example is thus expressed as $\mathcal{X}^i = [\mathbf{X}_c^i; \mathbf{X}_p^i] \in \mathbb{R}^{4 \times \mathbf{D}}$.
Given a dataset of $N$ labeled samples $\{(\mathcal{X}^i, \mathbf{Y}^i)\}_{i=1}^{N}$, where $\mathbf{Y}^i \in \{0,1\}^K$ is a one-hot vector denoting one of $K$ fluid categories, the goal is to learn a classification function as:
\begin{equation}
\mathcal{G}: \mathbb{R}^{4 \times \mathbf{D}} 
\rightarrow \{0,1\}^K,
\end{equation}
that predicts the fluid type $\mathbf{Y}$ from the multimodal 
input $\mathcal{X}$. 
In probabilistic terms, $\mathcal{G}$ aims to approximate the conditional distribution $p(\mathbf{Y}|\mathcal{X})$, enabling identification of fluids under varying flow conditions and sensor noise.
Figure~\ref{fig:mosaic-physical-attentionbased} illustrates 
the overview of the proposed PRIMS method.

\subsection{Physics-based Token Vectorization}
In the first stage of PRIMS, we convert raw Coriolis and pressure signals into compact and physically meaningful representations. 
Given two sensing modalities, the Coriolis mass flow $\mathbf{X}_c = [C_1; C_2]$ and the pressure $\mathbf{X}_p = [P_u; P_d]$, the signals are processed using physics-based transformations that align with fluid dynamic principles. 
This aims to reduce the model's reliance on purely data-driven feature extraction and to embed physical interpretability into the learned representations.

\vspace{5pt}
\noindent \textbf{Coriolis Physics Model.}
The Coriolis mass flow sensor contains a vibrating microchannel that converts fluid motion into measurable electrical signals~\cite{mi4010022}. 
Two electrodes positioned symmetrically along the channel detect oscillatory signals $C_1(t)$ and $C_2(t)$. According to \cite{ALVERINGH2023114762}, the phase shift $\Delta\phi$ and resonance frequency $f_0$ are proportional to the mass flow rate $\dot{m}$ and fluid density $\rho$:
\begin{equation}
    \Delta\phi \propto \dot{m}, \qquad 
    f_0 \propto \frac{1}{2\pi}\sqrt{\frac{k_{\mathrm{eff}}}{m_c + V_f\rho}},
\end{equation}
where $k_{\mathrm{eff}}$ is the effective stiffness of the vibrating tube, $m_c$ is the tube mass, and $V_f$ is the internal fluid volume.
To extract these physical cues, the Short-Time Fourier Transform (STFT) is applied to $C_1$ and $C_2$ to obtain their spectra $\mathcal{F}_{1_k}$ and $\mathcal{F}_{2_k}$ for $0 \le k < F$. 
The corresponding magnitude and phase spectra are computed as:
\begin{equation}
    A_k = |\mathcal{F}_{1_k}| + |\mathcal{F}_{2_k}|, \quad
    \Delta \varphi_k = H_{\mathrm{notch}}(k)\left[\angle \mathcal{F}_{2_k} - \angle \mathcal{F}_{1_k}\right],
\end{equation}
where $H_{\mathrm{notch}}(k)$ is a notch filter~\cite{oppenheim2010dsp} used to suppress noise. 
The resulting Coriolis embedding is $Z_c = [A; \Delta\varphi] \in \mathbb{R}^{2\times F}$, which encodes frequency and phase information correlated with mass flow and density.

\vspace{5pt}
\noindent \textbf{Pressure Physics Model.}
The pressure modality $\mathbf{X}_p = [P_u; P_d]$ provides static and dynamic information from upstream and downstream sensors. 
We compute the mean and differential pressure features as:
\begin{equation}
   \mu_P = (P_u + P_d)/2, \qquad \Delta P = P_u - P_d, 
\end{equation}
yielding the representation $Z_p = [\mu_P; \Delta P] \in \mathbb{R}^{2\times D}$. 
These features describe steady-state and transient pressure variations that are essential for flow and viscosity estimation.

\vspace{5pt}
\noindent \textbf{Embedding and Positional Encoding.}
The physics-derived features $(Z_c, Z_p)$ are projected through modality-specific Multi-layer Perceptrons into an embedding space of dimension $d_\text{embed}$ as:
\begin{equation}
    E_c = \text{ReLU}(Z_c W_c), \quad 
    E_p = \text{ReLU}(Z_p W_p),
\end{equation}
where $W_c$ and $W_p$ are learnable weights. 
To encode temporal dependencies, sinusoidal positional embeddings are added as:
\begin{equation}
    E'_c = E_c + \text{PE}, \quad E'_p = E_p + \text{PE},
\end{equation}
with $\text{PE}_{(pos,2i)} = \sin\left(\frac{pos}{10000^{2i/d_\text{embed}}}\right)$ and $\text{PE}_{(pos,2i+1)} = \cos\left(\frac{pos}{10000^{2i/d_\text{embed}}}\right)$.
The outputs $E'_c, E'_p \in \mathbb{R}^{2\times d_\text{embed}}$ are then used as input tokens for subsequent physics-based reasoning.

\subsection{Physical Component Synthesizer}
We design this module to model the interdependencies among flow, pressure, and density, and to extract viscosity-related representations that are both physically consistent and data-driven.
According to the Hagen--Poiseuille law~\cite{cengelFundamentalsThermalFluidSciences2001}, the pressure drop $\Delta P$ across a microchannel is proportional to the flow rate $\dot{m}$, modulated by viscosity and geometric parameters:  
\begin{equation}\label{eq:viscosity}
    \Delta P = \frac{8\eta L}{\rho \pi r_\text{eff}^4} \dot{m} = \frac{8\nu L}{\pi r_\text{eff}^4} \dot{m},
\end{equation}
where $\eta$ and $\nu$ denote the dynamic and kinematic viscosities, respectively, $L$ is the channel length, $\rho$ is the fluid density, and $r_\text{eff}$ is the effective radius.  
This relationship forms the foundation for learning viscosity representations within the model. 
We capture the dependencies among these quantities through an attention-based module that estimates kinematic and dynamic viscosity in a physics-based approach.

\vspace{5pt}
\noindent \textbf{Kinematic Viscosity Module.}
We model kinematic viscosity by jointly encoding Coriolis and pressure representations through bidirectional cross-attention.
Given embeddings $E'_c$ and $E'_p$, two Multi-Head Attention (MHA) blocks establish cross-modal alignment:
\begin{equation}
\begin{split}
    h_{\mathcal{V}_{1}} &= \text{MHA}(E'_p,\; E'_c,\; E'_c), \\
    h_{\mathcal{V}_{2}} &= \text{MHA}(E'_c,\; E'_p,\; E'_p),
\end{split}
\end{equation}
where $\text{MHA}(X_Q, X_K, X_V)$ follows the standard formulation of~\cite{vaswani2017attention}.
This bidirectional design allows pressure features to attend to flow and density cues while Coriolis features attend to pressure dynamics, encouraging the learned kinematic viscosity representations to respect the physical coupling between modalities.
We use two separate attention blocks rather than a single cross-attention because the Hagen–Poiseuille relation expresses kinematic viscosity as the ratio $\nu \propto \Delta P/\dot{m}$, where the pressure drop $\Delta P$ and the mass flow $\dot{m}$ contribute equally. A single cross-attention would make one modality the query and reduce the other to context, breaking this balance, whereas two blocks let each modality query the other in turn, giving both an equal role in the kinematic viscosity representation.
The two cross-attention outputs are then fused via a point-wise $1\!\times\!1$ convolution that reduces the channel dimension:
\begin{equation}
    h_{\mathcal{V}} = \text{Conv}_{1 \times 1}
    \big([h_{\mathcal{V}_{1}};\; h_{\mathcal{V}_{2}}]\big) 
    \in \mathbb{R}^{2 \times d_{\text{embed}}},
\end{equation}
where $[h_{\mathcal{V}_{1}};\; h_{\mathcal{V}_{2}}] \in \mathbb{R}^{4 \times d_{\text{embed}}}$ denotes the channel-wise concatenation of the two cross-attention outputs 
and the $1\!\times\!1$ convolution applies a learnable linear projection $W_{\text{conv}} \in \mathbb{R}^{2 \times 4}$ independently at each embedding position, reducing four channels back to two.

\vspace{5pt}
\noindent \textbf{Dynamic Viscosity Module.}
Since dynamic viscosity relates to kinematic viscosity scaled 
by density, we derive its representation by attending the fused kinematic viscosity embedding to density-related features via cross-attention:
\begin{equation}
    h_{\eta} = \text{MHA}(h_{\mathcal{V}},\; E'_c,\; E'_c),
\end{equation}
where $E'_c$ carries density-related information from the Coriolis modality.
By conditioning on $E'_c$ as keys and values, the module learns to modulate the kinematic viscosity representation with density cues, producing a dynamic viscosity feature that reflects the underlying physical coupling between the two quantities.


\subsection{Physics-guided Fusion}
To unify all physics-informed representations for fluid classification, we introduce the Physics-Guided Fusion (PGF) module based on multi-head self-attention.
This mechanism jointly processes embeddings of flow, pressure, density, and viscosity, enabling token-level interactions that capture cross-physical dependencies in a shared latent space:
\begin{align}
    S_\text{all} &= [E'_c;\; E'_p;\; h_{\mathcal{V}_1};\;  h_{\mathcal{V}_2};\; h_{\eta}] \in \mathbb{R}^{10 \times d_\text{embed}}, \\
    S_\text{MHA} &= \text{MHA}(S_\text{all},\; S_\text{all},\; S_\text{all}).
\end{align}
We retain the two directional outputs $h_{\mathcal{V}_1}$ and $h_{\mathcal{V}_2}$ here, rather than their fused form $h_{\mathcal{V}}$, so that self-attention can weigh each coupling direction independently per fluid. 
The fused $h_{\mathcal{V}}$ is instead optimized for the dynamic viscosity synthesis step, which could introduce unexpected biases.
The attended representations are then refined through a residual 
connection with the Adaptive Spatial Sync (ASS) 
module~\cite{nguyen2025imot}, which recalibrates channel-wise 
features through spatially adaptive gating, followed by layer 
normalization:
\begin{align}
    S' &= \text{LayerNorm}\!\big(S_\text{all} + \text{ASS}(S_\text{MHA})\big).
\end{align}
A position-wise feed-forward network (FFN) with GeLU activation is subsequently applied to enrich the per-token representations:
\begin{align}
    \text{FFN}(S') &= W_2\;\text{GeLU}(W_1 S' + b_1) + b_2,
\end{align}
where $W_1 \in \mathbb{R}^{d_\text{embed} \times d_\text{ff}}$, $W_2 \in \mathbb{R}^{d_\text{ff} \times d_\text{embed}}$, and $b_1, b_2$ are learnable biases.
A second residual connection with ASS and layer normalization yields the final fused output:
\begin{align}
    S_\text{out} &= \text{LayerNorm}\!\big(S' + \text{ASS}(\text{FFN}(S'))\big).
\end{align}
The output representation $S_\text{out} \in \mathbb{R}^{10 \times d_\text{embed}}$, serves as a universal, physics-integrated summary of the entire signal ensemble.


\subsection{Classification Head}
The fused representation from the PGF module is mapped to the 
target fluid category through a lightweight classification head.
We aggregate across all modality-channel tokens via global 
average pooling:
\begin{equation}
    \bar{s} = \frac{1}{M}\sum_{i=1}^{M} S_{\text{out}}[i] 
    \;\in\; \mathbb{R}^{d_{\text{embed}}},
\end{equation}
where $M$ corresponds to the total number of tokens. 
The pooled vector is then passed through two fully connected layers with dropout regularization. 
The network is trained by minimizing the cross-entropy 
loss:
\begin{equation}
    \mathcal{L}_{\text{CE}} = -\sum_{j=1}^{C} y_j 
    \log \hat{y}_j,
\end{equation}
where $y$ and $\hat{y}$ denote the ground-truth and 
predicted probabilities, respectively.

\section{Evaluation}
\label{sec:experiments}
\subsection{Experimental Conditions}
\noindent \textbf{Measuring Conditions.}
We obtain the data under controlled variations in environmental 
and flow conditions.
Temperature is varied from \SI{288}{\kelvin} to \SI{308}{\kelvin} in \SI{5}{\kelvin} increments, pressure from \SI{4}{\bar} to \SI{6}{\bar} in \SI{0.5}{\bar} steps, and mass flow from \SI{0}{\gram\per\hour} to \SI{5}{\gram\per\hour} with increments between \SI{0.3}{\gram\per\hour} and \SI{1}{\gram\per\hour}.
We conduct experiments with five representative fluids: Nitrogen (N\textsubscript{2}), Water (H\textsubscript{2}O), Isopropanol (IPA), Ethanol (EtOH), and Acetone (Ace).
A total of 5{,}540 measurement streams are acquired, each containing four synchronized sensor signals (two Coriolis electrodes and two pressure channels) sampled at \SI{10}{\kilo\hertz} for one second. 
Each stream is divided into 50 non-overlapping segments of 
\SI{20}{\milli\second}, yielding 277{,}000 samples in total.
The key physical properties of the five fluids are summarized in Table~\ref{tab:fluids}.

\vspace{5pt}
\noindent \textbf{Comparison Methods and Evaluation Metric.}
We compare PRIMS with representative baselines for fluid classification, including physics-based analytical modeling and data-driven architectures such as ConvNet and BiLSTM introduced in~\cite{ALVERINGH2023114762}.
For attention-based models, we conduct experiments with state-of-the-art time-series models, including TimeXer~\cite{wang2024timexer}, iTransformer~\cite{ICLR2024_2ea18fdc}, and TQNet~\cite{lin2025temporal}.
The task is formulated as a multi-class classification problem. 
Performance is evaluated using the macro-averaged F1-score over stratified 10-fold cross-validation, where each fold preserves the class distribution of the full dataset.
To prevent data leakage, the stratified 10-fold cross-validation is performed at the acquisition condition level. All segments recorded under the same temperature, pressure, and flow-rate combination are assigned to the same fold.

\vspace{5pt} 
\noindent \textbf{Model and Hyperparameters.}
We set the attention dimension to 88, corresponding to the 
selected frequency cutoff bins, and use 8 attention heads. 
The classification head consists of two fully connected layers 
with a hidden dimension of 64. We optimize the model using 
stochastic gradient descent with a learning rate of $10^{-3}$, 
a batch size of 8, and train for 50 epochs. All baseline models 
are implemented following their original configurations.

\subsection{Results}

\begin{table}[t]
\centering
\caption{Comparison of F1 scores (\%) for each fluid type and average of PRIMS and all baseline methods.
All methods are evaluated across all mass flow rates at a 10\,kS/s sampling rate, with results averaged over 10-fold cross-validation.
For TimeXer, iTransformer, TQNet, and PRIMS, per-class and macro-averaged F1 (\%), reported as mean$_{\pm\text{std}}$ over 5 random seeds.
Results for Analytical, ConvNet, and BiLSTM are taken from~\cite{ALVERINGH2023114762}.
Best mean per column is highlighted in \textbf{bold}.}
\label{tab:sota-indist-comparison}
\small
\setlength{\tabcolsep}{1.5pt}
\resizebox{\linewidth}{!}{
\begin{tabular}{lccccccc}
\toprule
\textbf{Method} & \textbf{Params} & \textbf{N\textsubscript{2}} & \textbf{H\textsubscript{2}O} & \textbf{IPA} & \textbf{EtOH} & \textbf{Ace} & \textbf{Average} \\
\midrule
Analytical~\cite{ALVERINGH2023114762} & --- & 25.00 & 100.0 & 79.11 & 64.00 & 88.08 & 71.16 \\
ConvNet~\cite{ALVERINGH2023114762} & 0.007 M & 15.00 & 46.00 & 40.00 & 44.00 & 68.00 & 42.60 \\
BiLSTM~\cite{ALVERINGH2023114762} & 0.085 M & \textbf{100.0} & 74.00 & 77.00 & 85.00 & 86.00 & 84.40 \\
TimeXer~\cite{wang2024timexer}  & \hspace{2pt} 8.53 M & $\mathbf{100.00}_{\pm0.00}$ & $\mathbf{100.00}_{\pm0.00}$ & $38.97_{\pm3.71}$ & $99.41_{\pm0.16}$ & $67.91_{\pm2.71}$ & $81.26_{\pm0.29}$ \\
iTransformer~\cite{ICLR2024_2ea18fdc} & \hspace{2pt} 6.42 M & \hspace{5pt} $99.99_{\pm0.01}$ & $\mathbf{100.00}_{\pm0.01}$ & $92.11_{\pm1.37}$ & $\mathbf{99.94}_{\pm0.06}$ & $94.84_{\pm0.77}$ & $97.37_{\pm0.44}$ \\
TQNet~\cite{lin2025temporal} & \hspace{2pt} 0.79 M  & $\mathbf{100.00}_{\pm0.00}$ & $\mathbf{100.00}_{\pm0.00}$ & $49.51_{\pm3.19}$ & $99.86_{\pm0.10}$ & $69.33_{\pm0.99}$ & $83.74_{\pm0.66}$ \\
\midrule
PRIMS (Ours) & \hspace{2pt} 0.46 M &  $99.91_{\pm0.06}$ & \hspace{5pt} $99.97_{\pm0.02}$ & $\mathbf{97.04}_{\pm0.09}$ & $99.86_{\pm0.10}$ & $\mathbf{97.81}_{\pm0.06}$ & $\mathbf{98.92}_{\pm0.03}$ \\
\bottomrule
\end{tabular}
}
\end{table}


\vspace{5pt}
\noindent \textbf{Comparison with Baselines.}
As shown in 
Table~\ref{tab:sota-indist-comparison}, 
PRIMS achieves the highest overall performance across all baselines, with an average F1 of 98.92\%. 
The purely physics-based analytical method yields an average of 71.16\%, highlighting the difficulty of modeling nonlinear sensor interactions using handcrafted equations alone. 
By embedding physical relationships within the attention mechanism, PRIMS effectively distinguishes fluids with similar properties, such as IPA, EtOH, and Ace, where purely data-driven ConvNet and BiLSTM methods~\cite{ALVERINGH2023114762} exhibit substantial performance degradation.
Moreover, PRIMS significantly outperforms TimeXer~\cite{wang2024timexer} and TQNet~\cite{lin2025temporal} in average score, surpassing them by 17.66\% and 15.18\%, respectively. 
Compared to iTransformer~\cite{ICLR2024_2ea18fdc}, which achieves 97.37\% with 6.42 million trainable parameters, PRIMS attains superior performance with only 0.46 million parameters, a 14$\times$ reduction in model size.
In addition, PRIMS is the most stable across random seeds: its average F1 varies by only $\pm0.03\%$, roughly an order of magnitude smaller than TimeXer ($\pm0.29\%$), iTransformer ($\pm0.44\%$), and TQNet ($\pm0.66\%$).
These results confirm that integrating physics-based priors into multimodal attention substantially enhances both accuracy and robustness in fluid classification.

\begin{table*}[t]
\centering
\caption{
Ablation study of PRIMS under different module configurations. PTV = Physics-based Token Vectorization; PCS = Physical Component Synthesizer; PGF = Physics-Guided Fusion.
All experiments are conducted at a sampling rate of 10\,kS/s across all flow rates.
Best results are highlighted in \textbf{bold}.
}
\label{tab:ablation-study-remove-block}
\small
{
\begin{tabular}{ccccccccccc}
\toprule
\textbf{Row} & \textbf{PTV} & \textbf{PCS} & \textbf{PGF} & \textbf{Model Size} & \textbf{N\textsubscript{2}} & \textbf{H\textsubscript{2}O} & \textbf{IPA} & \textbf{EtOH} & \textbf{Ace} & \textbf{Average} \\
\midrule
1 & \checkmark & \checkmark & \checkmark & {0.46 M} & \textbf{100.0} & \textbf{100.0} & \textbf{97.20} & \textbf{99.82} & \textbf{97.80} & \textbf{98.96} \\ 
2 & \checkmark &  & \checkmark & 0.35 M & 99.92 & 99.97 & 93.70 & 98.90 & 94.59 & 97.42 \\ 
3 & \checkmark & \checkmark &  & 0.16 M & \textbf{100.0} & \textbf{100.0} & 92.22 & 99.04 & 90.45 & 96.34 \\ 
4 & \checkmark &  &  & 0.05 M & 59.79 & 66.65 & 92.21 & 39.22 & 54.29 & 58.16 \\ 
5 &  & \checkmark & \checkmark & {2.17 M} & \textbf{100.0} & \textbf{100.0} & 93.46 & \textbf{100.0} & 95.56 & 97.80 \\ 
\bottomrule
\end{tabular}
}
\end{table*}

\vspace{5pt}
\noindent \textbf{Ablation Study.}
Table~\ref{tab:ablation-study-remove-block} presents the contribution of each proposed component. 
Removing any module consistently degrades performance, confirming the complementary role of all three design stages.
Excluding the Physical Component Synthesizer (PCS) leads to 
a reduction of 1.54\% in average F1-score (Row~1~$\rightarrow$~Row~2). 
Although this drop is small, PCS provides physically grounded viscosity representations that improve interpretability and contribute to the robustness observed in the out-of-distribution experiments.
The absence of the Physics-Guided Fusion (PGF) module causes a larger drop of 2.62\% (Row~1~$\rightarrow$~Row~3), highlighting its essential role in integrating cross-physical relationships.
When both PCS and PGF are removed, the average F1-score drops sharply to 58.16\% (Row~1~$\rightarrow$~Row~4), demonstrating that the physics-informed synthesis and fusion stages are critical for achieving strong classification performance beyond what tokenization alone can provide.
Finally, removing the Physics-based Token Vectorization (PTV) while keeping both PCS and PGF (Row~1~$\rightarrow$~Row~5) lowers the average F1-score by only 1.16\%, yet enlarges the model from 0.46\,M to 2.17\,M parameters (about 4.7$\times$).
This shows that PTV can compresses the raw signal into a compact, physically meaningful token representation.

\begin{table}[t]
\centering
\caption{
Out-of-distribution generalization under isolated 
temperature testing.
Each block holds out a single temperature 
as the test set; all remaining temperatures are used for training. 
Best results are highlighted in \textbf{bold}.
}
\label{tab:sota-temp-comparison}
\small
\setlength{\tabcolsep}{5pt}
{
\begin{tabular}{llcccccc}
\toprule
\textbf{Temp} & \textbf{Model} & \textbf{N\textsubscript{2}} & \textbf{H\textsubscript{2}O} & \textbf{IPA} & \textbf{EtOH} & \textbf{Ace} & \textbf{Average} \\
\midrule
288 K & TimeXer~\cite{wang2024timexer} & \textbf{100.0} & \textbf{100.0} & 3.02 & \textbf{91.58} & 67.55 & 72.43 \\
 & iTransformer~\cite{ICLR2024_2ea18fdc} & \textbf{100.0} & \textbf{100.0} & 65.60 & 74.45 & 41.15 & 76.24 \\
 & TQNet~\cite{lin2025temporal} & \textbf{100.0} & \textbf{100.0} & 27.91 & 77.31 & 47.54 & 70.55 \\
 & PRIMS (Ours) & 96.82 & 99.74 & \textbf{69.43} & 77.05 & \textbf{57.36} & \textbf{80.08} \\

\midrule

 293 K  & TimeXer~\cite{wang2024timexer} & \textbf{100.0} & \textbf{100.0} & 21.14 & 98.69 & 69.05 & 77.78 \\
 & iTransformer~\cite{ICLR2024_2ea18fdc} & \textbf{100.0} & \textbf{100.0} & 49.71 & \textbf{99.83} & 11.01 & 72.11 \\
 & TQNet~\cite{lin2025temporal} & \textbf{100.0} & \textbf{100.0} & 48.00 & 97.96 & 54.57 & 80.11 \\
 & PRIMS (Ours) & \textbf{100.0} & 99.94 & \textbf{73.50} & 96.45 & \textbf{69.47} & \textbf{87.87} \\

\midrule

298 K & TimeXer~\cite{wang2024timexer} & \textbf{100.0} & \textbf{100.0} & 15.12 & 99.14 & 68.23 & 76.50 \\
 & iTransformer~\cite{ICLR2024_2ea18fdc} & \textbf{100.0} & \textbf{100.0} & 66.26 & 84.32 & 44.06 & 78.93 \\
 & TQNet~\cite{lin2025temporal} & \textbf{100.0} & \textbf{100.0} & 26.16 & \textbf{99.78} & 64.53 & 78.09 \\
 & PRIMS (Ours) & 99.95 & 99.98 & \textbf{93.63} & 98.17 & \textbf{92.81} & \textbf{96.91} \\

\midrule

303 K & TimeXer~\cite{wang2024timexer} & \textbf{100.0} & \textbf{100.0} & 31.67 & 99.13 & 65.28 & 79.22 \\
 & iTransformer~\cite{ICLR2024_2ea18fdc} & \textbf{100.0} & \textbf{100.0} & 81.07 & 99.86 & 81.69 & 92.52 \\
 & TQNet~\cite{lin2025temporal} & \textbf{100.0} & \textbf{100.0} & 43.02 & \textbf{100.0} & 62.09 & 81.02 \\
 & PRIMS (Ours) & 99.64 & \textbf{100.0} & \textbf{99.64} & 99.82 & \textbf{100.0} & \textbf{99.82} \\

\midrule
308 K & TimeXer~\cite{wang2024timexer} & 99.75 & 99.20 & 53.01 & 97.18 & 36.56 & 77.14 \\
 & iTransformer~\cite{ICLR2024_2ea18fdc} & 68.13 & 95.88 & 65.92 & 90.97 & \textbf{84.29} & 81.04 \\
 & TQNet~\cite{lin2025temporal} & \textbf{100.0} & \textbf{99.94} & 53.35 & 99.23 & 62.35 & 82.97 \\
 & PRIMS (Ours) & 99.32 & 98.35 & \textbf{69.95} & \textbf{97.77} & 72.90 & \textbf{87.66} \\
\bottomrule
\end{tabular}
}
\end{table}

\vspace{5pt}
\noindent \textbf{Out-of-Distribution Generalization.}
Table~\ref{tab:sota-temp-comparison} evaluates all models under out-of-distribution conditions, where an entire temperature regime is held out during training. 
PRIMS achieves the highest average F1-score across all five temperature splits, ranging from 80.08\% at 288\,K to 99.82\% at 303\,K. 
The advantage is most pronounced on fluids with similar rheological profiles: at 298\,K, PRIMS achieves 93.63\% on IPA while iTransformer~\cite{ICLR2024_2ea18fdc} and TQNet~\cite{lin2025temporal} reach only 66.26\% and 26.16\%, respectively. 
This suggests that the physics-aware representations capture temperature-invariant features rather than memorizing regime-specific patterns. 

\begin{figure}[t]
\centering
\setlength{\tabcolsep}{2pt}
\begin{tabular}{@{}cccc@{}}
 & {\small Dynamic viscosity} & 
   {\small Kinematic viscosity 1} & 
   {\small Kinematic viscosity 2} \\[2pt]
\raisebox{-.5\height}{\rotatebox{90}{\small {Fold 1}}} &
\raisebox{-.5\height}{\includegraphics[width=0.30\textwidth]{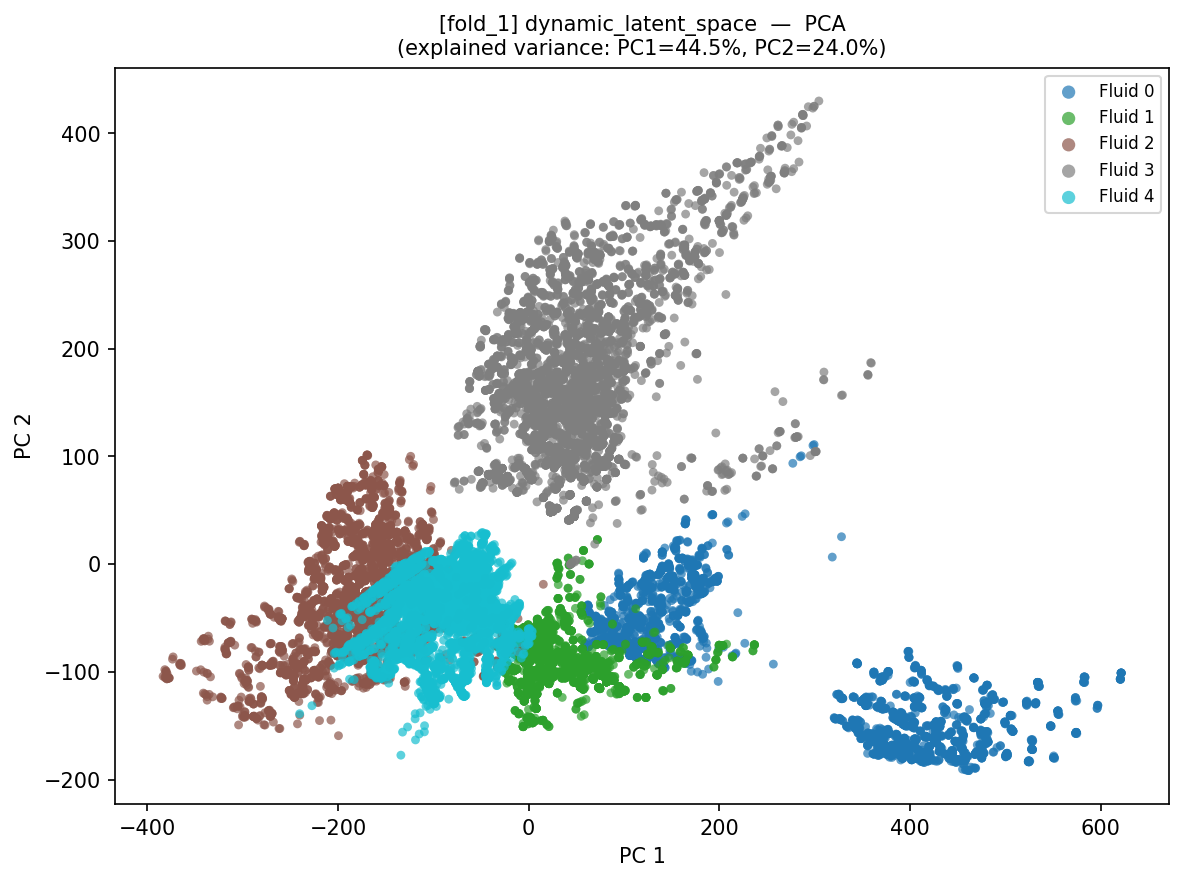}} &
\raisebox{-.5\height}{\includegraphics[width=0.30\textwidth]{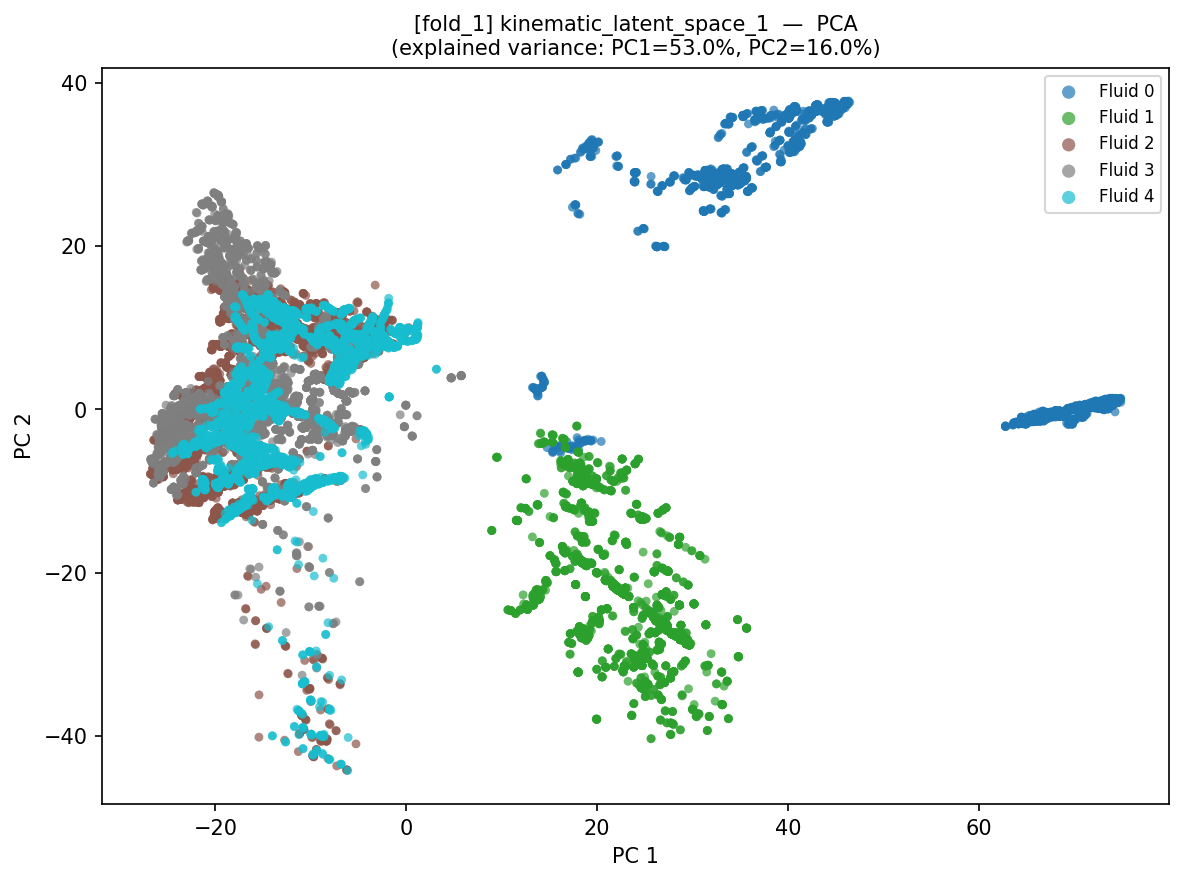}} &
\raisebox{-.5\height}{\includegraphics[width=0.30\textwidth]{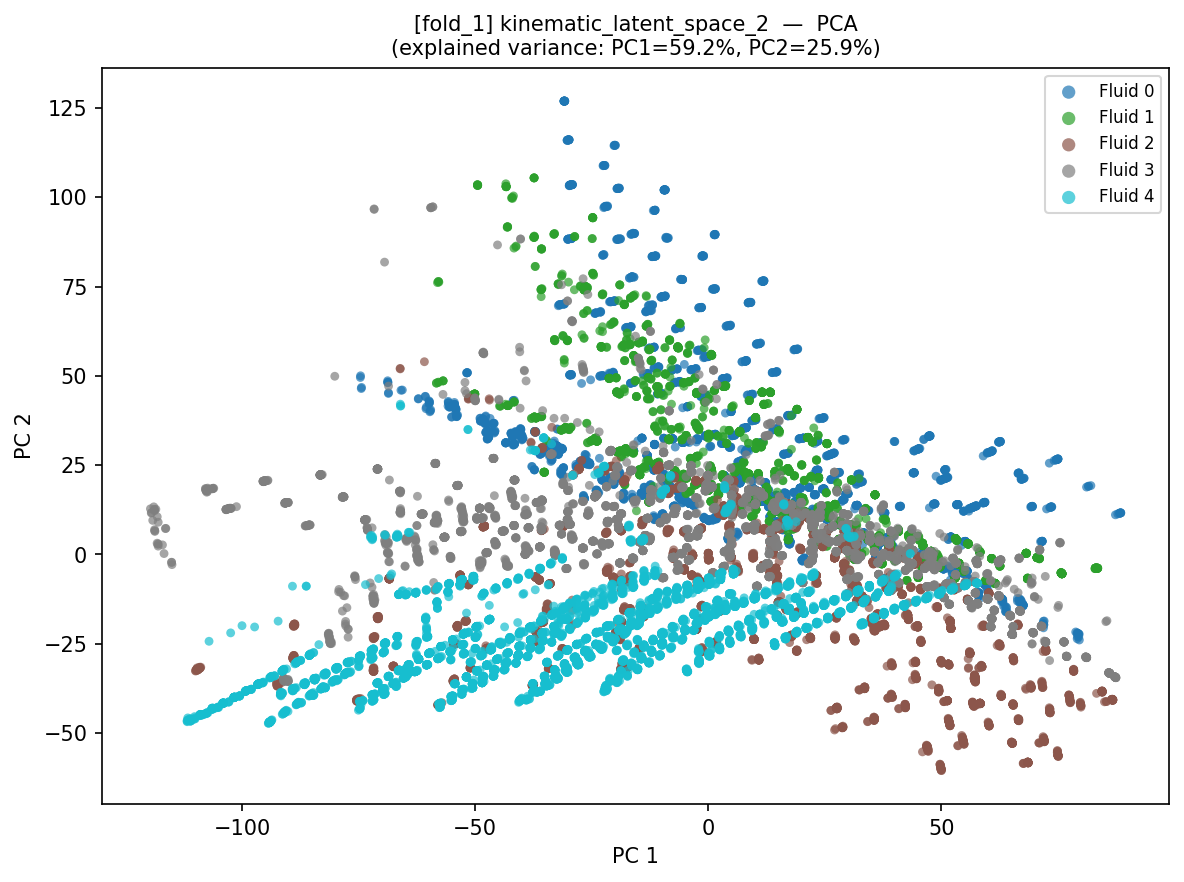}} \\[4pt]
\raisebox{-.5\height}{\rotatebox{90}{\small {Fold 2}}} &
\raisebox{-.5\height}{\includegraphics[width=0.30\textwidth]{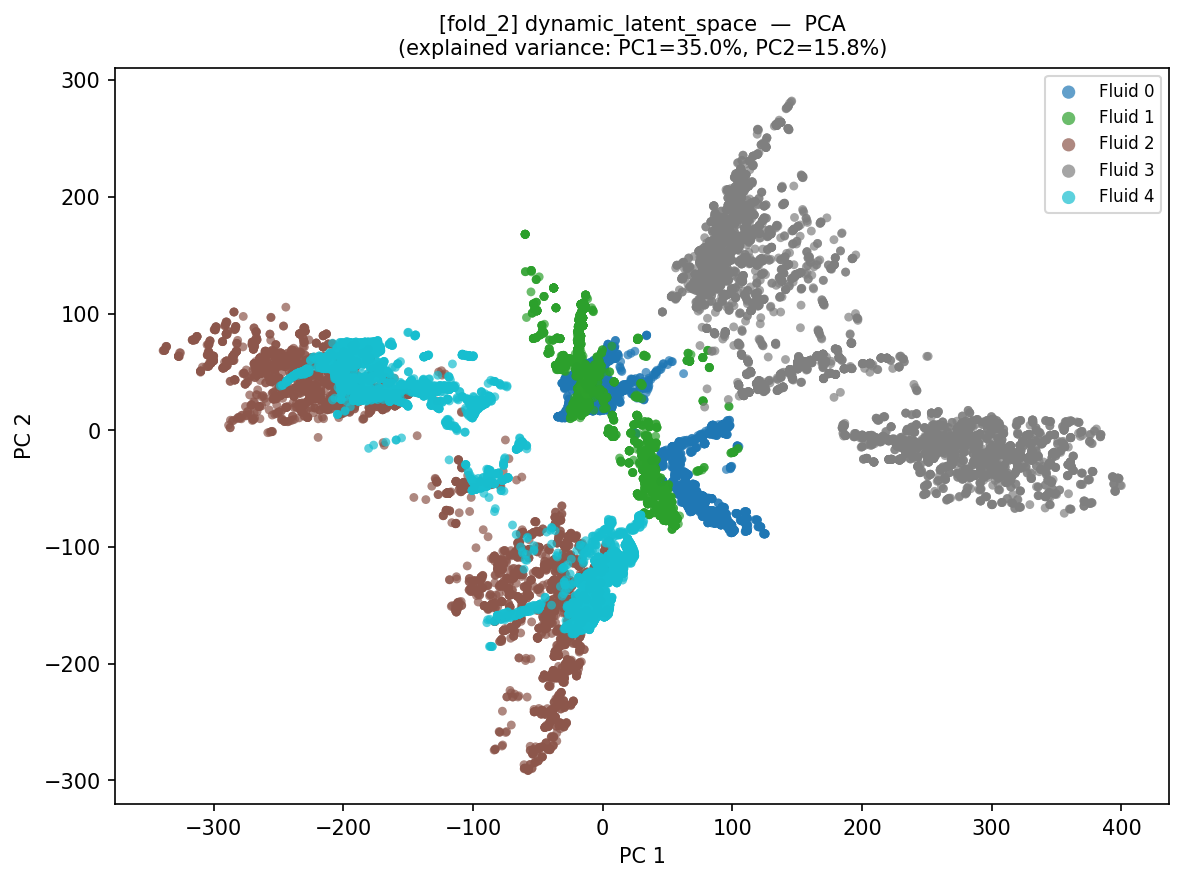}} &
\raisebox{-.5\height}{\includegraphics[width=0.30\textwidth]{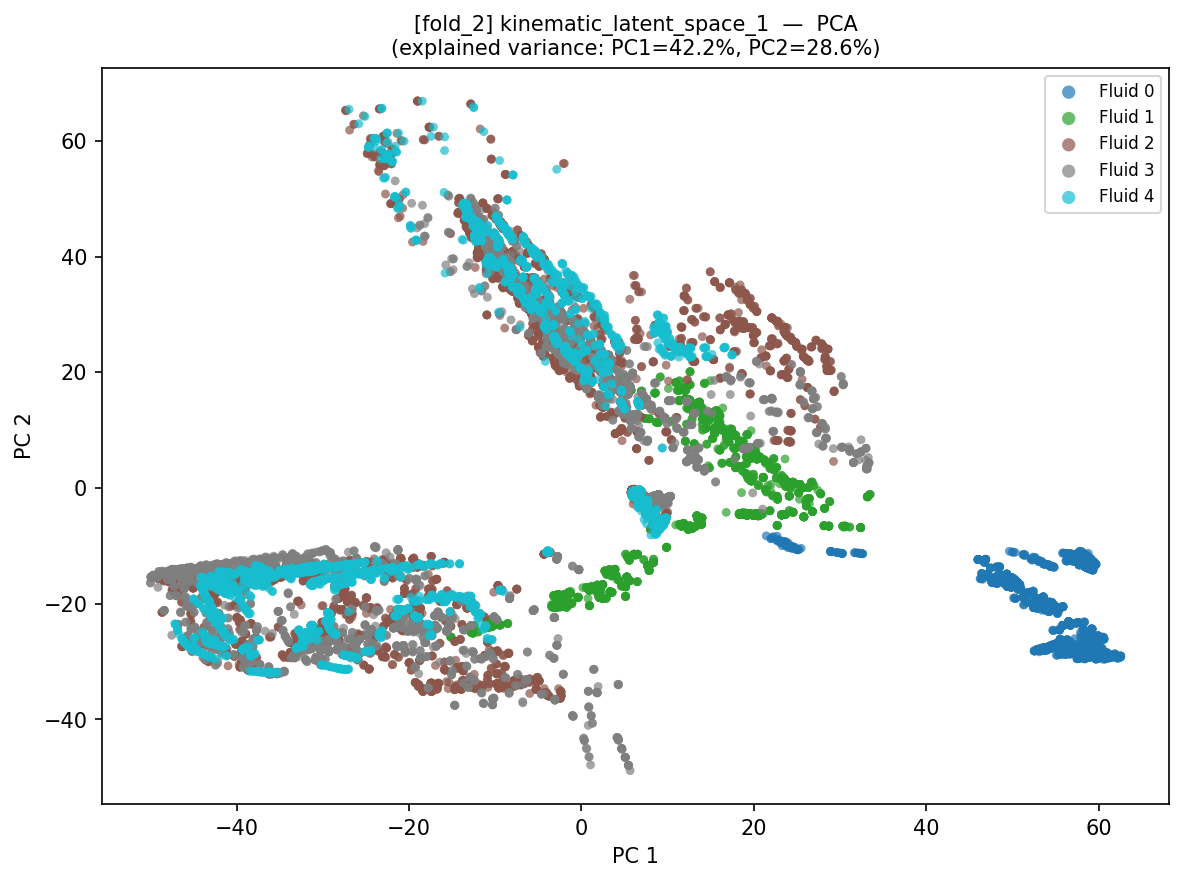}} &
\raisebox{-.5\height}{\includegraphics[width=0.30\textwidth]{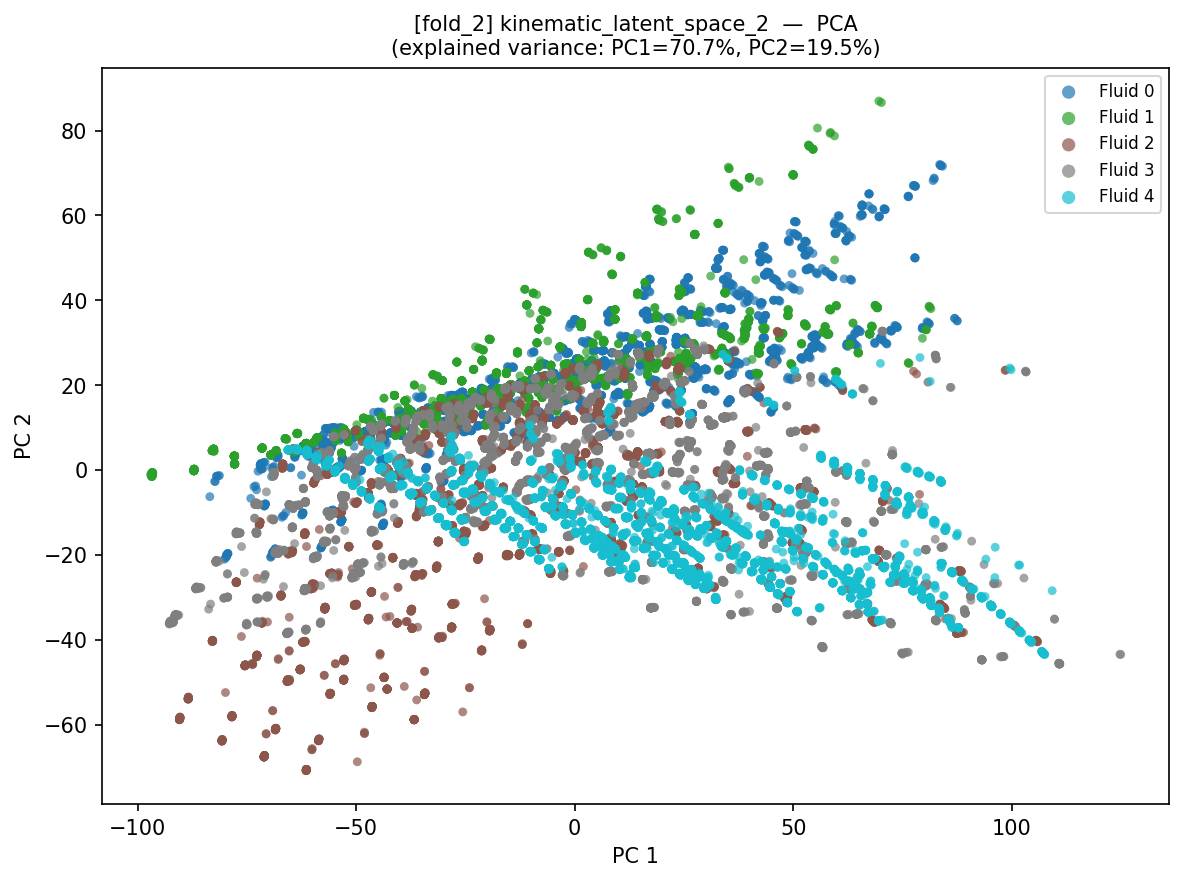}} \\
\end{tabular}
\caption{
PCA visualization of the PRIMS latent spaces across two cross-validation folds. 
Each point represents a sample colored by fluid type (0: N\textsubscript{2}, 1: H\textsubscript{2}O, 2: IPA, 3: EtOH, 4: Ace). 
The dynamic viscosity space (left column) shows clear cluster separation, kinematic viscosity~1 (middle) shows partial separability, and kinematic viscosity~2 (right) reveals considerable overlap. 
The consistent geometry across both folds indicates stable, physically grounded representations.
}
\label{fig:latent_space_pca}
\end{figure}

\subsection{Latent Space Analysis}
Figure~\ref{fig:latent_space_pca} visualizes the learned latent spaces of the three viscosity branches via PCA. 
The dynamic viscosity space (left column) produces the clearest separation: all five fluid classes form distinct clusters in both folds, confirming that this branch learns highly discriminative features aligned with the underlying rheological differences among fluids. 
Kinematic viscosity~1 (middle column) reveals a complementary pattern: N\textsubscript{2} and H\textsubscript{2}O are well isolated, while IPA, EtOH, and Acetone exhibit partial overlap, consistent with their similar kinematic viscosity values in Table~\ref{tab:fluids}. 
Kinematic viscosity~2 (right column) shows considerable overlap among fluid classes, suggesting that this branch captures shared flow characteristics rather than class-specific features.
Crucially, the cluster geometry remains consistent across Fold~1 and Fold~2, where different data partitions are used for training. 
This consistency demonstrates that the learned representations reflect stable physical structure rather than artifacts of a particular data split, corroborating the robust out-of-distribution generalization observed in Table~\ref{tab:sota-temp-comparison}.

\begin{table}[t]
  \caption{Computational profiling of PRIMS and baseline models 
    using Open Neural Network Exchange (ONNX) Runtime on a single CPU core.
  Columns report model size (MB), parameter count, operation count (Ops), 
  inference latency (ms), and throughput in samples per second (S/s). 
  Best results are highlighted in \textbf{bold}.
    }
  \label{tab:profiling_summary}
  \centering\small
  \setlength{\tabcolsep}{2.5pt}
  \begin{tabular}{lrrrrr}
    \toprule
    \textbf{Model}
      & \textbf{Size (MB)}
      & \textbf{Params}
      & \textbf{Ops}
      & \textbf{Latency (ms)}
      & \textbf{Throughput (S/s)} \\
    \midrule
    TimeXer~\cite{wang2024timexer}         
      & 41.97 & 8{,}525{,}370  
      & 511 & 3.54 & 283 \\
    iTransformer~\cite{ICLR2024_2ea18fdc}    
      & 24.52 & 6{,}418{,}437  
      & 203 & 0.50 & 2{,}017 \\
    TQNet~\cite{lin2025temporal}           
      & 3.03 & 790{,}149  
      & \textbf{121} & \textbf{0.11} 
      & \textbf{9{,}525} \\
    PRIMS (Ours) 
      & \textbf{1.68} & \textbf{463{,}507} 
      & 135 & 0.18 & 5{,}698 \\
    \bottomrule
  \end{tabular}
\end{table}

\subsection{Computational Efficiency}
To fairly compare computational cost across different frameworks, we export all models to the Open Neural Network Exchange (ONNX) format and profile them using ONNX Runtime on a single CPU core. 
It provides a framework-independent representation that ensures consistent measurement of model size, operator count, inference latency, and throughput across all methods.
Table~\ref{tab:profiling_summary} compares the computational 
profiles of Transformer-based methods. 
PRIMS is the most compact, requiring only 1.68\,MB and 463{,}507 parameters, which represents an 14$\times$ reduction in size compared to iTransformer~\cite{ICLR2024_2ea18fdc} and a 25$\times$ reduction compared to TimeXer~\cite{wang2024timexer}. 
While TQNet~\cite{lin2025temporal} achieves the lowest latency 
of 0.11\,ms and highest throughput of 9{,}525\,S/s, its 
classification accuracy of 83.29\% average F1 falls substantially 
behind PRIMS at 98.96\%.
PRIMS achieves a competitive latency of 0.18\,ms with 5{,}698\,S/s throughput, offering the most favorable accuracy-efficiency trade-off among all compared methods. 
These results confirm that embedding physics-based inductive biases not only improves classification performance but also reduces model complexity, as the architecture can rely on physically meaningful representations.

\section{Conclusion}
\label{sec:conclusion}
In this study, we proposed PRIMS, a physics-aware multimodal Transformer for fluid classification in microfluidic systems. 
PRIMS integrates physics-driven tokenization, cross-attention-based viscosity synthesis, and attention-based fusion to align data-driven learning with underlying fluid dynamics. 
Evaluations on a five-fluid benchmark demonstrate that PRIMS achieves 98.96\% average F1-score with only 0.46 million parameters, a 14$\times$ reduction compared to state-of-the-art Transformer-based methods while maintaining strong generalization. 
These results confirm that embedding physical priors directly into the model architecture improves both accuracy and efficiency, enabling practical deployment on resource-constrained devices.

\vspace{5pt}
\noindent \textbf{Limitations and Future Work.} 
PRIMS currently embeds physical knowledge through architectural design alone, without explicit auxiliary losses to enforce relationships. 
Incorporating physics-based constraint losses, such as those derived from the Hagen--Poiseuille law, could further regularize the latent space toward physically consistent representations. 
Additionally, the Hagen--Poiseuille relation assumes laminar, Newtonian, and incompressible flow; while this approximation is justified under low-flow, moderate-pressure operating regime, it may introduce modelling error for gases at higher flow rates where compressibility becomes significant.
Future work will integrate explicit physical constraint losses, extend the framework to compressible and non-Newtonian flows, and validate on broader fluid vocabularies and real-world conditions.

\section*{Acknowledgements}
\label{sec:acknowledgements}
This publication is part of the project MOSAIC: enhancement of MicrOfluidic Sensing with deep symbolic Artificial IntelligenCe with file number 19985 of the research programme Open Technology Programme which is (partly) financed by the Dutch Research Council (NWO).

\bibliographystyle{splncs04}
\bibliography{mybibliography}

@inproceedings{hu2025physics,
  author    = {Chen, Huaguan and Liu, Yang and Sun, Hao},
  title     = {PINP: Physics-Informed Neural Predictor with latent estimation of fluid flows},
  booktitle = {International Conference on Learning Representations},
  year      = {2025},
}

@ARTICLE{4675306,
  author={Smith, Richard and Sparks, Douglas R. and Riley, Diane and Najafi, Nader},
  journal={IEEE Transactions on Industrial Electronics}, 
  title={A MEMS-Based Coriolis Mass Flow Sensor for Industrial Applications}, 
  year={2009},
  volume={56},
  number={4},
  pages={1066-1071},
}

@article{haneveld_jmm_2010,
  author  = {Haneveld, J. and Lammerink, T.S.J. and de Boer, M.J. and Sanders, R.G.P. and Mehendale, S. and L{\"o}tters, J.C. and Dijkstra, M. and Wiegerink, R.J.},
  title   = {Modeling, design, fabrication and characterization of a micro Coriolis mass flow sensor},
  journal = {Journal of Micromechanics and Microengineering},
  volume  = {20},
  number  = {12},
  pages   = {125001},
  year    = {2010},
}

@misc{bronkhorst_ml120v21_datasheet,
  title   = {mini CORI-FLOW ML120V21 Datasheet},
  author  = {{Bronkhorst High-Tech B.V.}},
  year    = {2016},
  note    = {Accessed: 2025-10-04}
}

@article{raissi_jcp_2019,
  author  = {Raissi, Maziar and Perdikaris, Paris and Karniadakis, George E.},
  title   = {Physics-informed neural networks: A deep learning framework for solving forward and inverse problems involving nonlinear partial differential equations},
  journal = {Journal of Computational Physics},
  volume  = {378},
  pages   = {686--707},
  year    = {2019},
}

@book{kirby_cup_2010,
  title={Micro-and nanoscale fluid mechanics: transport in microfluidic devices},
  author={Kirby, Brian J},
  year={2010},
  publisher={Cambridge university press}
}

@article{koupai2022self,
  title={Self-supervised multimodal fusion transformer for passive activity recognition},
  author={Koupai, Armand K and Bocus, Mohammud J and Santos-Rodriguez, Raul and Piechocki, Robert J and McConville, Ryan},
  journal={IET Wireless Sensor Systems},
  volume={12},
  number={5-6},
  pages={149--160},
  year={2022},
  publisher={Wiley Online Library}
}

@article{yang_entropy_2022,
  author  = {Yang, Ye and Lu, Jiangang},
  title   = {A Fusion Transformer for Multivariable Time Series Forecasting: The Mooney Viscosity Prediction Case},
  journal = {Entropy},
  volume  = {24},
  number  = {4},
  pages   = {528},
  year    = {2022},
}

@inproceedings{zhou2021informer,
  title={Informer: Beyond efficient transformer for long sequence time-series forecasting},
  author={Zhou, Haoyi and Zhang, Shanghang and Peng, Jieqi and Zhang, Shuai and Li, Jianxin and Xiong, Hui and Zhang, Wancai},
  booktitle={Proceedings of the AAAI conference on artificial intelligence},
  volume={35},
  number={12},
  pages={11106--11115},
  year={2021}
}

@article{li2022efficientformer,
  title={Efficientformer: Vision transformers at mobilenet speed},
  author={Li, Yanyu and Yuan, Geng and Wen, Yang and Hu, Ju and Evangelidis, Georgios and Tulyakov, Sergey and Wang, Yanzhi and Ren, Jian},
  journal={Advances in Neural Information Processing Systems},
  volume={35},
  pages={12934--12949},
  year={2022}
}

@inproceedings{edgevit_eccv_2022,
  title={Edgevits: Competing light-weight cnns on mobile devices with vision transformers},
  author={Pan, Junting and Bulat, Adrian and Tan, Fuwen and Zhu, Xiatian and Dudziak, Lukasz and Li, Hongsheng and Tzimiropoulos, Georgios and Martinez, Brais},
  booktitle={European conference on computer vision},
  pages={294--311},
  year={2022},
  organization={Springer}
}

@article{ALVERINGH2023114762,
title = {Fluid classification with integrated flow and pressure sensors using machine learning},
journal = {Sensors and Actuators A: Physical},
volume = {363},
pages = {114762},
year = {2023},
issn = {0924-4247},
author = {D. Alveringh and D.V. Le and J. Groenesteijn and J. Schmitz and J.C. Lötters}
}

@inproceedings{nguyen2025imot,
  title={iMoT: Inertial Motion Transformer for Inertial Navigation},
  author={Nguyen, Son Minh and Le, Duc Viet and Havinga, Paul},
  booktitle={Proceedings of the AAAI Conference on Artificial Intelligence},
  volume={39},
  number={6},
  pages={6209--6217},
  year={2025}
}

@article{ejeian2019design,
  title={Design and applications of MEMS flow sensors: A review},
  author={Ejeian, Fatemeh and Azadi, Shohreh and Razmjou, Amir and Orooji, Yasin and Kottapalli, Ajay and Warkiani, Majid Ebrahimi and Asadnia, Mohsen},
  journal={Sensors and Actuators A: Physical},
  volume={295},
  pages={483--502},
  year={2019},
  publisher={Elsevier}
}

@article{van1995multi,
  title={Multi-parameter detection in fluid flows},
  author={Van Kuijk, J and Lammerink, TSJ and De Bree, H-E and Elwenspoek, M and Fluitman, JHJ},
  journal={Sensors and actuators A: Physical},
  volume={47},
  number={1-3},
  pages={369--372},
  year={1995},
  publisher={Elsevier}
}

@inproceedings{zubavicius2024fluid,
  title={Fluid Viscosity and Density Determination With Machine Learning-Enhanced Coriolis Mass Flow Sensors},
  author={Zubavicius, Romas and Alveringh, Dennis and Poel, Mannes and Groenesteijn, Jarno and Sanders, Remco GP and Wiegerink, Remco J and L{\"o}tters, Joost C},
  booktitle={2024 IEEE 37th International Conference on Micro Electro Mechanical Systems},
  pages={82--85},
  year={2024},
}

@inproceedings{schut2018fully,
  title={Fully integrated mass flow, pressure, density and viscosity sensor for both liquids and gases},
  author={Schut, Thomas Victor Paul and Alveringh, Dennis and Sparreboom, Wouter and Groenesteijn, Jarno and Wiegerink, Remco J and L{\"o}tters, J C},
  booktitle={2018 IEEE Micro Electro Mechanical Systems},
  pages={218--221},
  year={2018},
}

@article{cengelFundamentalsThermalFluidSciences2001,
  title = {{{Fundamentals of }}{{{{Thermal-Fluid Sciences}}}}},
  author = {Cengel, Ya and Turner, Rh and Smith, R},
  year = {2001},
  month = nov,
  journal = {Applied Mechanics Reviews},
  volume = {54},
  number = {6},
  pages = {B110-B112},
  issn = {0003-6900, 2379-0407},
  urldate = {2024-11-19},
  langid = {english},
}

@article{vaswani2017attention,
  title={Attention is all you need},
  author={Vaswani, Ashish and Shazeer, Noam and Parmar, Niki and Uszkoreit, Jakob and Jones, Llion and Gomez, Aidan N and Kaiser, {\L}ukasz and Polosukhin, Illia},
  journal={Advances in neural information processing systems},
  volume={30},
  year={2017}
}

@Article{mi4010022,
AUTHOR = {Sparreboom, Wouter and Van de  Geest, Jan and Katerberg, Marcel and Postma, Ferry and Haneveld, Jeroen and Groenesteijn, Jarno and Lammerink, Theo and Wiegerink, Remco and Lötters, Joost},
TITLE = {Compact Mass Flow Meter Based on a Micro Coriolis Flow Sensor},
JOURNAL = {Micromachines},
VOLUME = {4},
YEAR = {2013},
NUMBER = {1},
PAGES = {22--33},
ISSN = {2072-666X},
}

@article{mustafa2023machine,
  title={Machine learning based microfluidic sensing device for viscosity measurements},
  author={Mustafa, Adil and Haider, Daniyal and Barua, Arnab and Tanyeri, Melikhan and Erten, Ahmet and Yalcin, Ozlem},
  journal={Sensors \& Diagnostics},
  volume={2},
  number={6},
  pages={1509--1520},
  year={2023},
  publisher={Royal Society of Chemistry}
}

@article{yang2025integrated,
  title={An Integrated Microfluidic Microwave Array Sensor with Machine Learning for Enrichment and Detection of Mixed Biological Solution},
  author={Yang, Sen and Wang, Yanxiong and Jiang, Yanfeng and Qiang, Tian},
  journal={Biosensors},
  volume={15},
  number={1},
  pages={45},
  year={2025},
  publisher={MDPI}
}

@article{park2024machine,
  title={Machine learning-driven innovations in microfluidics},
  author={Park, Jinseok and Kim, Yang Woo and Jeon, Hee-Jae},
  journal={Biosensors},
  volume={14},
  number={12},
  pages={613},
  year={2024},
  publisher={MDPI}
}

@article{wu2021autoformer,
  title={Autoformer: Decomposition transformers with auto-correlation for long-term series forecasting},
  author={Wu, Haixu and Xu, Jiehui and Wang, Jianmin and Long, Mingsheng},
  journal={Advances in neural information processing systems},
  volume={34},
  pages={22419--22430},
  year={2021}
}

@book{oppenheim2010dsp,
  title     = {Discrete-Time Signal Processing},
  author    = {Oppenheim, Alan V. and Schafer, Ronald W.},
  edition   = {3},
  year      = {2010},
  publisher = {Pearson},
  address   = {Upper Saddle River, NJ}
}

@article{wang2024timexer,
  title={Timexer: Empowering transformers for time series forecasting with exogenous variables},
  author={Wang, Yuxuan and Wu, Haixu and Dong, Jiaxiang and Qin, Guo and Zhang, Haoran and Liu, Yong and Qiu, Yunzhong and Wang, Jianmin and Long, Mingsheng},
  journal={Advances in Neural Information Processing Systems},
  volume={37},
  pages={469--498},
  year={2024}
}

@inproceedings{ICLR2024_2ea18fdc,
 author = {Liu, Yong and Hu, Tengge and Zhang, Haoran and Wu, Haixu and Wang, Shiyu and Ma, Lintao and Long, Mingsheng},
 booktitle = {International Conference on Representation Learning},
 editor = {B. Kim and Y. Yue and S. Chaudhuri and K. Fragkiadaki and M. Khan and Y. Sun},
 pages = {11116--11140},
 title = {iTransformer: Inverted Transformers Are Effective for Time Series Forecasting},
 volume = {2024},
 year = {2024}
}

@inproceedings{
lin2025temporal,
title={Temporal Query Network for Efficient Multivariate Time Series Forecasting},
author={Shengsheng Lin and Haojun Chen and Haijie Wu and Chunyun Qiu and Weiwei Lin},
booktitle={Forty-second International Conference on Machine Learning},
year={2025},
}

@INPROCEEDINGS{7994261,
  author={Alveringh, D. and Schut, T. V. P. and Wiegerink, R. J. and Sparreboom, W and Lötters, J. C.},
  booktitle={2017 19th International Conference on Solid-State Sensors, Actuators and Microsystems}, 
  title={Resistive pressure sensors integrated with a coriolis mass flow sensor}, 
  year={2017},
  volume={},
  number={},
  pages={1167-1170},
  }

@inproceedings{nguyen2025multisensor,
  title={MultiSensor-Home: A wide-area multi-modal multi-view dataset for action recognition and Transformer-based sensor fusion},
  author={Nguyen, Trung Thanh and Kawanishi, Yasutomo and John, Vijay and Komamizu, Takahiro and Ide, Ichiro},
  booktitle={Proceedings of the 19th IEEE International Conference on Automatic Face and Gesture Recognition},
  pages={1--10},
  year={2025},
}

@inproceedings{pham2026padm,
  title={{PADM}: A Physics-aware Diffusion Model for Attenuation Correction},
  author={Pham, Trung Kien and Vu, Hoang Minh and Chu, Anh Duc and Nguyen, Dac Thai and Nguyen, Trung Thanh and Truong, Thao Nguyen and Son, Mai Hong and Nguyen, Thanh Trung and Le Nguyen, Phi},
  booktitle={Proceedings of the 2026 IEEE/CVF Winter Conference on Applications of Computer Vision},
  pages={2606--2615},
  year={2026}
}
\end{document}